\documentclass[aps,prb,groupaddress,twocolumn,nopacs,letterpaper,nobibnotes]{revtex4}

\usepackage{color}
\usepackage{amsmath}
\usepackage{graphics}
\usepackage{graphicx,xcolor}
\usepackage{txfonts}
\usepackage{epsfig}
\usepackage{fancyhdr}
\usepackage[pdfstartview=FitH,colorlinks=true,linkcolor=blue,citecolor=blue,urlcolor=blue]{hyperref}

\renewcommand{\vec}[1]{\bf #1 \rm}
\newcommand{\q}{\vec{q}}
\newcommand{\up}{\uparrow}
\newcommand{\down}{\downarrow}

\begin{document}

\title{Spin-Imbalanced Pairing and Fermi Surface Deformation in Flat Bands}

\author{Kukka-Emilia Huhtinen}
\thanks{These authors contributed equally to this work.}

\author{Marek Tylutki}
\thanks{These authors contributed equally to this work.}

\author{Pramod Kumar}
\author{Tuomas I.\ Vanhala}
\author{Sebastiano Peotta}
\author{P\"{a}ivi T\"{o}rm\"{a}}
\email{paivi.torma@aalto.fi}

\affiliation{Department of Applied Physics, Aalto University, 00076
  Aalto, Finland}

\date{\today}

\begin{abstract}
  We study the attractive Hubbard model with spin imbalance on two
  lattices featuring a flat band: the Lieb and kagome lattices. We
  present mean-field phase diagrams featuring exotic superfluid
  phases, similar to the Fulde-Ferrell-Larkin-Ovchinnikov (FFLO)
  state, whose stability is confirmed by dynamical mean-field theory
  (DMFT). The nature of the pairing is found to be richer than
    just the Fermi surface shift responsible for the usual FFLO
  state. The presence of a flat band allows for changes in the
  particle momentum distributions at null energy cost. This
  facilitates formation of nontrivial superfluid phases via multiband
  Cooper pair formation: the momentum distribution of the spin
  component in the flat band deforms to mimic the Fermi surface of the
  other spin component residing in a dispersive band. The Fermi
  surface of the unpaired particles that are typical for gapless
  superfluids becomes deformed as well. The results highlight the
  profound effect of flat dispersions on Fermi surface instabilities,
  and provide a potential route for observing spin-imbalanced
  superfluidity and supercondutivity.
\end{abstract}

\maketitle

\section{Introduction}
Interactions in fermion systems may cause Fermi surface (FS)
instabilities, for instance towards pairing~\cite{Cooper1956} or
symmetry-breaking deformations of the FS, called the Pomeranchuk
instability (PI)~\cite{Pomeranchuk1959}. These mechanisms lead to
various phases of matter such as both conventional and high-$T_c$
superconductivity~\cite{Bednorz1986,Lee2006,Bloch2008}, topological
phases thereof~\cite{Sato2017}, the two superfluid phases of different
symmetry in $^3$He~\cite{Leggett1975} or superfluidity in lattice
systems of ultracold fermions predicted by the Hubbard
model~\cite{Torma2015}, including models with spin-orbit
coupling~\cite{Iskin2013,Qu2013}. In the repulsive Hubbard model, the
superfluidity may coexist with the magnetic stripe
order~\cite{Vanhala2017}, or with PI as in
Refs.~\cite{Kiesel2013,Kitatani2017}. Spin-imbalanced superfluidity,
on the other hand, has been predicted to simultaneously display
pairing, superfluidity, and gapless excitations (FSs). These exotic
phases of matter spontaneously break symmetries of the system, for
instance rotational or translational, in addition to the breaking of
the $U(1)$ gauge symmetry characteristic of any BCS-type
superfluid. In the FFLO state~\cite{Fulde1964,Larkin1965} the Cooper
pairs carry a finite momentum. Deformed FS superfludity
(DFS)~\cite{Muether2002,Sedrakian2005} has been proposed as another
alternative that gives a lower energy than the conventional BCS
theory. Such predictions have remained elusive, supported only by
indirect experimental evidence~\cite{Casalbuoni2004,Liao2010}. Phase
separation, instead of exotic spin-imbalanced superfluids, has been
observed in ultracold quantum
gases~\cite{Zwierlein2006s,Partridge2006,Partridge2006a,Shin2008,Nascimbene2009};
this is consistent with predictions for continuum
systems~\cite{Sheehy2006,Sheehy2007,Recati2008}, although theory
suggests that lattice systems may stabilize the FFLO state due to
nesting~\cite{Koponen2007,Wolak2012,Baarsma2016,Kinnunen2017,Cichy2017,Heikkinen2014}. In
general, singularities in the density of states (DOS) are known to
enhance FS instabilities. Here we show that multiband lattice systems
which possess the ultimate DOS singularity, namely a flat (constant)
energy band, allow deformations of the particle momentum distribution
without energy cost and thereby stabilize a new type of
spin-imbalanced superfluidity. We find that the origin of the pairing
is different from a simple minority particle FS shift conventionally
responsible for FFLO states.

\begin{figure}
\includegraphics[width=.99\columnwidth]{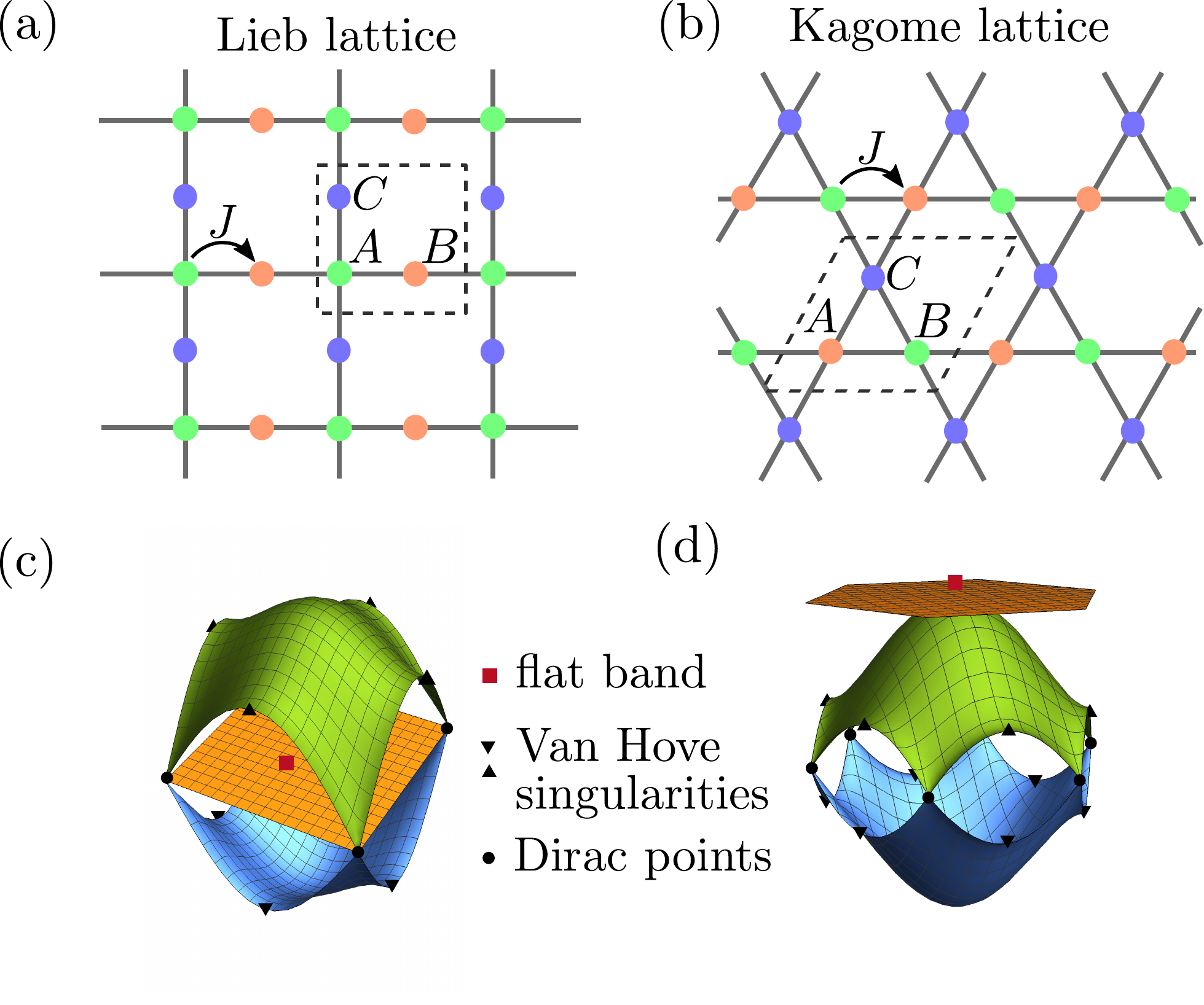}\\
\caption{Two lattice geometries featuring a FB: (a) a Lieb
  lattice and (b) a kagome lattice. The elementary cells are delimited
  with dashed lines.
  Three sites that constitute an elementary cell
  are labeled as A, B and C. Below, single-particle band structures of
  these lattices: (c) Lieb and (d) kagome with two dispersive bands
  and one FB. The singularities and Dirac points are shown in Fig.~\ref{fig.phase_diagrams} with lines.}
\label{fig.schematic}
\end{figure}

\section{Model}
We study two examples of a Hubbard model with a flat band (FB) in the
single-particle energy spectrum: a Lieb lattice and a kagome lattice
Hubbard model. Both lattices have three sublattices and feature two
dispersive bands and a FB,
\begin{align}
&E_{\pm,{\rm Lb}}({\bf k})  = \pm J \sqrt2 \sqrt{2 +\cos k_x+\cos k_y}, && E_{\rm FB, Lb} = 0 \\
&E_{\pm,{\rm Kg}}({\bf k}) = J(1\pm\sqrt{3+2\Lambda({\bf k})}), && E_{\rm FB, Kg} = 2J, 
\end{align}
where
$\Lambda({\bf k}) = \sum_{i=1}^3
\cos({\bf k}\cdot {\bf a_i})$. 
The vectors $\bf{a_1}$ and $\bf{a_2}$ are the primitive
vectors of the kagome lattice, and $\bf{a_3} = \bf{a_1}-\bf{a_2}$.
The indices Lb and Kg refer to the Lieb and kagome lattices
respectively.  
By $J$, which we also use as the unit of energy, we denote the hopping
strength between the neighboring lattice sites. Hereafter the lattice
constant $a$ is assumed $a = 1$. 

The lattices and the band structures
are shown in Fig~\ref{fig.schematic}. Importantly, such lattices have
been experimentally realized for ultracold
gases~\cite{PhysRevLett.108.045305,Taiee1500854,PhysRevLett.118.175301},
in designer lattices made by atomistic
control~\cite{Slot2017,Drost2017}, in optical
analogues~\cite{PhysRevLett.114.245504,PhysRevLett.114.245503} and
also implementations with superconducting circuits have been proposed
theoretically~\cite{PhysRevB.93.054116}.  We choose to fix chemical
potentials and therefore consider the grand-canonical ensemble. The
real-space grand-canonical Hamiltonian reads
\begin{equation}
H = \sum_\sigma \sum_{\bf i \alpha, j \beta} \psi_{\bf i \alpha \sigma}^\dag \mathcal{H}_{\bf i \alpha, j \beta}^{\vphantom{\dag}} \psi_{\bf j \beta \sigma}^{\vphantom{\dag}} - \sum_{\sigma} \mu_\sigma N_\sigma + H_{\rm int} ~,
\label{eq.fullH}
\end{equation}
where the lattice information is contained in the single-particle
Hamiltonian $\mathcal{H}_{\bf i \alpha, j\beta}$ responsible for
hopping between the lattice sites, $\alpha$ and $\beta$ are the
sublattice (orbital) indices. In our model, we consider only nearest
neighbor hopping for both lattices. The particle number operator is
defined as
$N_\sigma = \sum_{\bf i \alpha} \psi_{\bf i \alpha \sigma}^\dag
\psi_{\bf i \alpha \sigma}^{\vphantom{\dag}}$,
and the on-site interaction enters as
$H_{\rm int} = U \sum_{\bf i \alpha} \psi_{\bf i \alpha \up}^\dag
\psi_{\bf i \alpha \down}^\dag \psi_{\bf i \alpha
  \down}^{\vphantom{\dag}} \psi_{\bf i \alpha
  \up}^{\vphantom{\dag}}$.
We define the average chemical potential as
$\mu = (\mu_\up + \mu_\down) / 2$ and the effective magnetic field as
$h = (\mu_\up - \mu_\down) / 2$.

The BCS (mean-field) approximation
of the Hamiltonian~(\ref{eq.fullH}) introduces a pairing field
$\Delta_{\bf i \alpha} = U\langle \psi_{\bf i \alpha \down} \psi_{\bf i
  \alpha \up} \rangle$, where the average denotes a ground state
expectation value at zero temperature and a grand-canonical average
at finite temperatures $k_{\rm B} T = 1 / \beta$.
We allow for an imbalance in chemical
potentials, $\mu_\up \neq \mu_\down$, so the particles in a Cooper pair
may have a nonzero center-of-mass momentum $\bf q$. This is reflected
by the Fulde-Ferrell (FF) ansatz for the pairing field, $\Delta_{\bf j
  \alpha} = \Delta_{\alpha} e^{i \bf q \cdot j}$. Since we assume our system
to be translationally invariant, we change the basis to the
quasi-momentum basis by performing a Fourier transform. 
After this transformation the mean-field Hamiltonian with the FF ansatz becomes
\begin{equation}
H_{\rm FF} = \sum_{\bf k}\, \left[  \Psi^\dag_{\bf k} \mathcal{H}_{\rm BdG}^{\vphantom{\dag}} \Psi_{\bf k}^{\vphantom{\dag}} - 3 \mu_\downarrow -  \frac1U {\rm Tr}\, \boldsymbol{\Delta}^\dag \boldsymbol{\Delta} \right] ~,
\end{equation}
where we introduced a Nambu spinor
$\Psi_{\bf k} = (c_{{\bf k}, A \up}^{\vphantom{\dag}}, c_{{\bf k}, B
  \up}^{\vphantom{\dag}}, c_{{\bf k}, C \up}^{\vphantom{\dag}},
c_{{\bf q - k}, A \down}^\dag, c_{{\bf q - k}, B \down}^\dag, c_{{\bf
    q - k}, C \down}^\dag)^T$
and the Bogoliubov-de-Gennes (BdG) Hamiltonian
\begin{equation}
\mathcal{H}_{\rm BdG} = \begin{pmatrix}
\mathcal{H}_{\bf k} - \mu_{\uparrow} & \boldsymbol{\Delta}\\
\boldsymbol{\Delta}^\dag & -\mathcal{H}_{\bf -k+q} + \mu_{\downarrow}\\
\end{pmatrix} ~.
\end{equation}
The pairing fields are collected into a diagonal matrix
$(\boldsymbol{\Delta})_{\alpha \beta} = \Delta_\alpha \delta_{\alpha \beta}$.

The single-particle Hamiltonian can be diagonalized as
$\mathcal{G}_{\bf k \sigma}^\dag \mathcal{H}_{\bf k \sigma}^{\vphantom{\dag}}
\mathcal{G}_{\bf k \sigma}^{\vphantom{\dag}} = {\bf \epsilon}_{\bf k \sigma}^{\vphantom{\dag}}$.
In this single-particle band basis, the field operators 
take the form
\begin{equation}
\begin{pmatrix} {\bf d}_{\bf k \up}^{\vphantom{\dag}}\\ {\bf d}_{\bf q - k \down}^\dag\\ \end{pmatrix} = \begin{pmatrix}
 \mathcal{G}_{\bf k \up}^\dag & 0\\
0 &  \mathcal{G}_{\bf q - k \down}^\dag\\
\end{pmatrix} \Psi_{\bf k} ,
\end{equation}
where the components of the collective vector 
$({\bf d}_{\bf k \up}^{\vphantom{\dag}}, {\bf d}_{\bf q - k
  \down}^\dag)^T$
correspond to different bands. A further unitary transformation to
quasi-particle basis,
$({\bf \gamma}_{\bf k, q \up}^{\vphantom{\dag}},{\bf \gamma}_{\bf k, q
  \down}^\dag)^T$,
diagonalizes the full BdG Hamiltonian, $\mathcal{H}_{\rm BdG}$. The
diagonalized Hamiltonian reads
\begin{equation}
H_{\rm FF} = \sum_{\bf k} (\gamma_{\bf k, q \up}^\dag {\bf E}_{\bf k,
  q \up}^{\vphantom{\dag}} \gamma_{\bf k, q \up}^{\vphantom{\dag}} +
\gamma_{\bf k, q \down}^\dag {\bf E}_{\bf k, q
  \down}^{\vphantom{\dag}} \gamma_{\bf k, q \down}^{\vphantom{\dag}})
+ \mathcal{E} ~,
\end{equation}
where ${\bf E}_{\bf k, q\, \sigma}$ are diagonal matrices of the
quasi-particle energies, and the energy offset
$\mathcal{E} = \sum_{\bf k} (-3 \mu_\downarrow + {\rm Tr}\,
\boldsymbol{\Delta}^\dag \boldsymbol{\Delta} / U - {\rm Tr}\, {\bf
  E}_{\bf k, q \down})$.
In order to find thermodynamically stable phases at finite
temperature, we look for global minima of the thermodynamic potential
$\Omega = -\ln \mathrm{Tr} \exp(-\beta H_{\rm FF}) / \beta$, which can
be calculated as
\begin{equation}
\Omega = -\frac{1}{\beta} \sum_{\bf{k}, \sigma}{\rm Tr} 
\ln [1 + \exp(-\beta\, {\bf E}_{\bf k, q\sigma})] + \mathcal{E} ~.
\end{equation}
We minimize it with respect to all components of $\boldsymbol{\Delta}$
and $\bf q$ independently.


\begin{figure}
\includegraphics[width=.99\columnwidth]{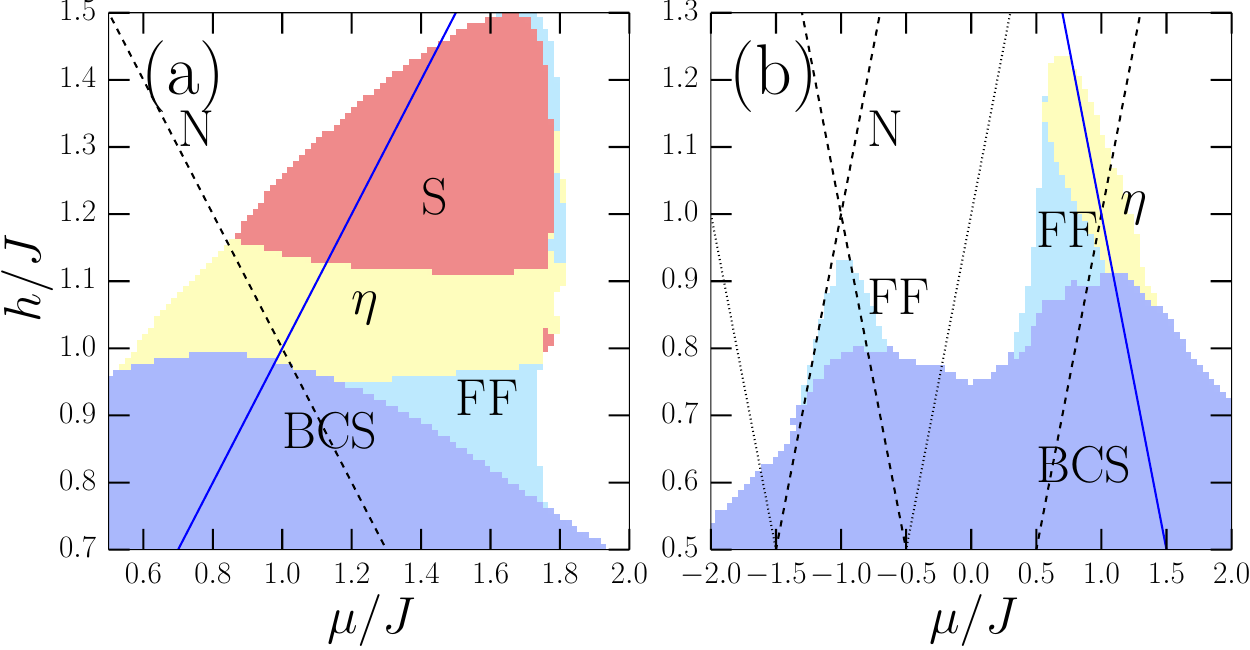}\\
\caption{Mean-field phase diagram for (a) a Lieb lattice and (b) kagome lattice
  at $U=-4J$ and $k_{\rm B}T=0.1J$. Here, $\mu=(\mu_\up+\mu_\down)/2$ and
  $h=(\mu_\up-\mu_\down)/2$. Dashed lines indicate Van Hove
  singularities, blue solid lines the FBs, and dotted lines the
  Dirac points. 
  They are determined as $\mu_{\up, \down} = E_{\rm s}$,
  i.e. where the chemical potential reaches the energy $E_{\rm s}$
  corresponding to a relevant point in the DOS.}
\label{fig.phase_diagrams}
\end{figure}

\begin{figure}
\includegraphics[width=.99\columnwidth]{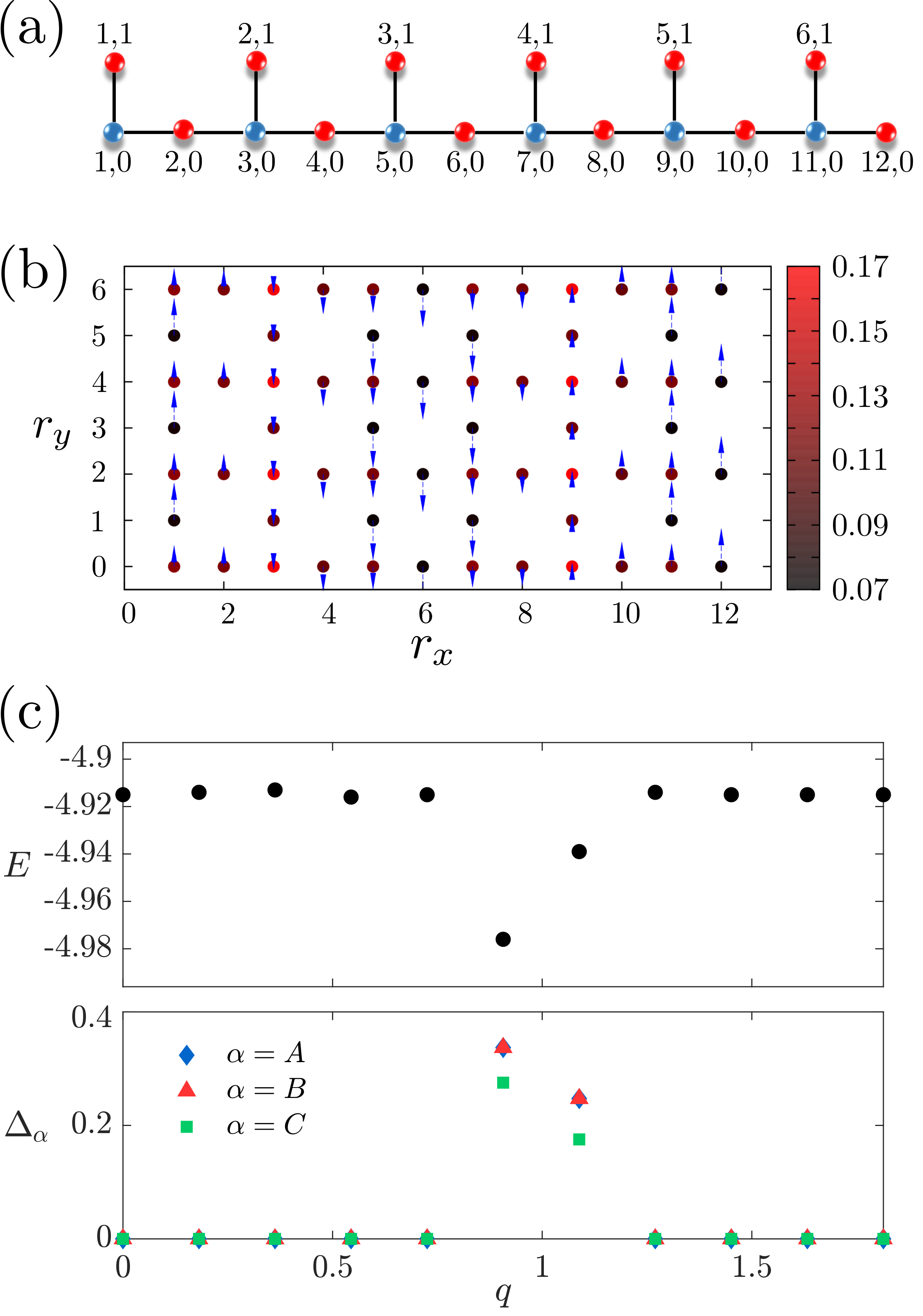}
\caption{(a) The 18 site cluster used for DMFT calculations for the
  Lieb lattice. (b) The s-wave order parameter, $\Delta(r_x,r_y)$
  (arrow length and direction), and spin-polarization
  $n_s(r_x,r_y)=n_{\uparrow}(r_x,r_y)-n_{\downarrow}(r_x,r_y)$ (dots
  with color scale) for different positions $(r_x,r_y)$ at
  $h\sim 1.40$, $ \mu \sim 0.0$, $U=-6J$ and $k_{\rm B}T=0.05 J$ in
  the FFLO state evaluated using DMFT for the Lieb lattice.  (c) Total
  energy per unit cell (upper panel) and order parameters (lower
  panel) computed using DMFT, as a function of the amplitude of
  ${\bf q}$ at lattice filling fractions $n_\up\approx 2.06$ and
  $n_\down\approx 1.62$ with $U=-4J$ and at zero temperature. In the
  lower panel, different symbols represent the order parameters in the
  three sites of the unit cell. In the FF state, two of the order
  parameters are equal due to the symmetry of the kagome lattice,
  while the third is smaller than the others due to the finite
  momentum ${\bf q}$, which breaks the symmetry of the lattice.}
\label{fig.dmft}
\end{figure}

\begin{figure*}
\includegraphics[width = .98\textwidth]{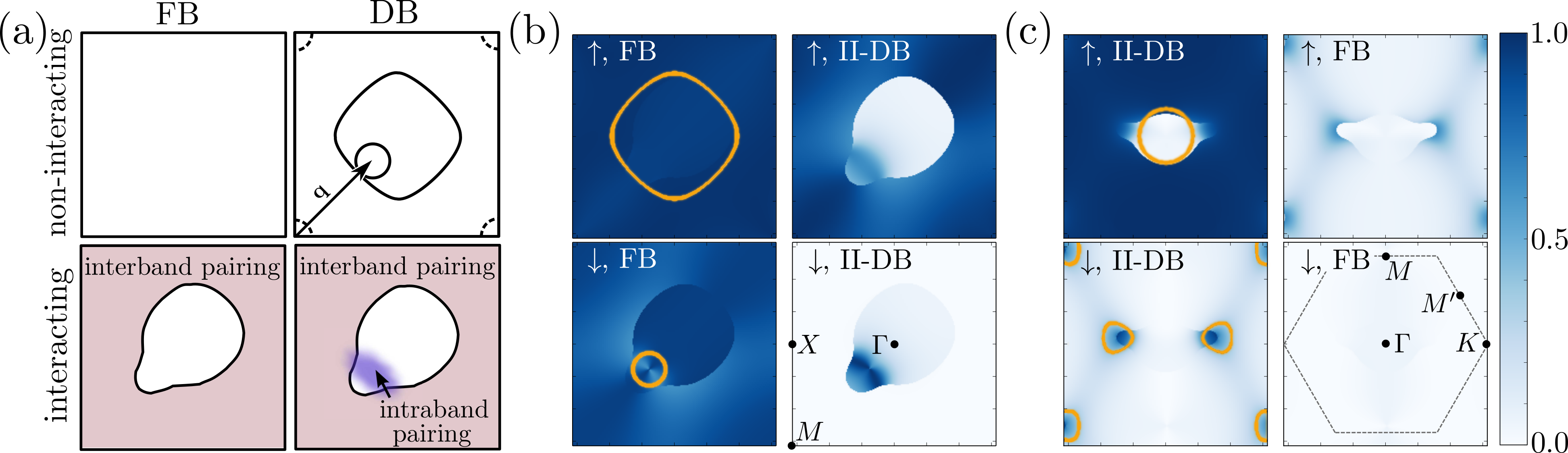}
\caption{(a) The FF pairing mechanism in a multiband system with a
  FB. The FS of the minority component shifts
  towards the FS of the majority component by the momentum
  $\q$. Both intra- and interband pairings (regions which are shown
  here in color) contribute to the overall effect. The calculated
  band-resolved density profiles $n_{{\bf k}n\sigma}$  in (b) the Lieb
  lattice for $\mu = 1.4 J$ and $h = 0.9 J$ and in (c) the kagome
  lattice for $\mu=0.6J$ and $h=1.1J$ demonstrate the discussed
  scenario. Orange lines indicate the noninteracting Fermi
  surfaces. High-symmetry points are marked, and dashed lines in (c)
  indicate the boundary of the Brillouin zone. 
}
\label{fig.band_densities}
\end{figure*}

\section{Phase diagrams}
We present mean field phase diagrams in Fig.~\ref{fig.phase_diagrams}
for both considered lattices, at interaction $U=-4J$ and temperature
$k_{\rm B}T=0.1J$.  Due to particle-hole symmetry, the phase diagram
for the Lieb lattice is symmetric with respect to the axis
$\mu=0$. This symmetry is absent in the kagome lattice.  We assume
$\mu_\up \geq \mu_\down$. In both cases, the BCS phase is favored for
sufficiently low chemical potential imbalance $h$.  As $h$ is
increased, the phase switches either to a normal phase, or to
nonuniform superfluidity with nonzero $\q$. We distinguish two such
phases: the FF and $\eta$ phase. In the FF region, ${\bf q}$ is in the
Brillouin zone (BZ) and grows until it reaches the boundary of the
BZ. There, it saturates at its maximum value at the $M$ point,
${\bf q}=(\pi,\pi)$ in the Lieb lattice and ${\bf q}=(0,\pi/\sqrt{3})$
in the kagome lattice. This means the order parameter oscillates with
a period equal to twice the lattice period. This phase, otherwise
similar to FFLO but with $\q$ having such a maximal value, is referred
to as the $\eta$ phase in the literature~\cite{Yang1989,Ptok2017}. In
the Lieb lattice, a third imbalanced superfluid phase with $\q=0$, the
so-called Sarma phase~\cite{Sarma1963,Liu2003,He2009,Kim2013}, is
found at large imbalance $h$. The focus of this article is on the FF
and $\eta$ phases, and the Sarma phase will be discussed in detail
in~\cite{TylutkiNew}.

In both lattices, we find that the DOS singularities are manifested in
the phase diagram. Nonuniform superfluidity occurs near crossing
points of singularity lines, where the density of states near the FSs of
both components is large. In the Lieb lattice, FB
singularities are always involved at interaction $U=-4J$. In the
kagome lattice, however, a smaller FF region is found away from the
FB, where the minority component reaches the Van Hove (VH)
singularity on the first dispersive band, and the majority component
reaches that on the second dispersive band.

Importantly, one can see from Fig.~\ref{fig.phase_diagrams} that the
FF and $\eta$ phases are stable mostly close to the flat band DOS
singularity. Near the flat band the FS of one component is small, or
even nonexistent, and one would expect pairing to be
suppressed. Indeed, in conventional BCS theory pairing is enhanced by
the size of the FS.  The formation of nonuniform superfluidity in our
case is not explained solely by matching of the FSs as in previous
literature~\cite{Kinnunen2017}, indicating there are other mechanisms
at play.

\section{Dynamical mean-field theory}
To verify the existence of the FF phase beyond the simple mean-field
approximation, we performed DMFT calculations in a partially real-space
formulation for both lattices.  Dynamical mean-field theory (DMFT)
maps a lattice problem to an effective single impurity problem taking
into account the lattice effects in a self-consistent manner. A
central quantity is the self-energy $\Sigma_{ij}(i \omega_n)$, where
$i$ and $j$ index the lattice sites and $\omega_n=\pi(2n+1)T$, where
$T$ is the temperature, are the fermionic Matsubara frequencies. Within
single-site DMFT the self-energy is assumed to be local to each site
$i$ and uniform over the whole lattice, so that
$\Sigma_{ij}(i \omega_n) \sim \delta_{ij}\Sigma(i \omega_n)$. For
inhomogeneous states such as the Fulde-Ferrell–-Larkin–-Ovchinnikov
phase (FFLO), however, the uniformity assumption breaks, as the order
parameter can be different for different lattice sites. To study such
states, we thus use a partially real-space cluster extension of
DMFT~\cite{PhysRevB.96.245127,RevModPhys.77.1027}, in which the
self-energy is still local but varies spatially for different sites in
the cluster, i.e. $\Sigma_{ij}(i\omega_n)=\Sigma_{i}(i\omega_n)\delta_{ij}$.

More rigorously, the DMFT method in Nambu-Gorkov formalism for a
given cluster can be described as follows. The local Green's
function of the lattice system limited to a single cluster can be
calculated as 
\begin{equation}
  \mathbf{G}(i\omega_n)= \frac{1}{N_{k}}\sum_{\mathbf{k}} 
  \left(
    \mathbf{G}^0(\mathbf{k},i\omega_n)^{-1}-\mathbf{\Sigma}(i\omega_n) 
  \right)^{-1}, \label{eq4}
\end{equation}
where the bold quantities are matrices whose dimension equals the
number of sites within the cluster and $N_k$ is the number of $k$-
points. Each component consists of a $(2 \times 2)$ matrix with normal
Green's functions as diagonal components, while the off-diagonal
components are anomalous Green's functions. Thus the $2\times 2$
block $\mathbf{G}(i \omega_n)_{ij}$ is the Green's function between
sites $i$ and $j$ of the cluster. The non-interacting Green's
function
$\mathbf{G}^0(\mathbf{k},i\omega_n)^{-1}_{ij}=
(i\omega_n+h)\delta_{ij}\sigma_0+\mu\delta_{ij}\sigma_z
-\mathbf{T}(\mathbf{k})_{ij}\sigma_z$, 
where $\mathbf{T}(\mathbf{k})$ is the superlattice Fourier transform
of the hopping matrix. The site diagonal self-energy at the $i$th site
is given by the following $(2 \times 2)$ matrix
$$
\mathbf{\Sigma}_i(i\omega_n)=
\begin{pmatrix} 
\Sigma_{i}(i\omega_n) & S_{i}(i\omega_n) \\
S_{i}(i\omega_n) & -\Sigma_{i}^{*}(i\omega_n)
\end{pmatrix}
$$
where $\Sigma(i\omega_n)$ ($S(i\omega_n)$) is the normal (anomalous)
part of the self-energy. For each site $i$ in the cluster, there is
an effective single impurity Anderson model, which is defined by the
dynamical Weiss mean-field
\begin{equation}
\mathbf{\mathcal{G}}_{i}(i\omega_n)^{-1}=
(\mathbf{G}(i\omega_n)_{ii})^{-1}+\mathbf{\Sigma}_{i}(i\omega_n).\label{eq5}
\end{equation}
Given the Weiss function $\mathbf{\mathcal{G}}_{i}$ for all $i$, we
calculate the self-energy of each of the impurity problems using a
continuous time quantum Monte-Carlo (CTINT)
algorithm~\cite{RevModPhys.83.349} for the Lieb lattice, and an exact
diagonalization (ED) solver for the kagome lattice. These new
self-energies are then used again in equation \ref{eq4} and the
process is iterated until a converged solution is found.  

For the Lieb lattice, the calculations were performed for a cluster of
18 sites, shown in Fig.~\ref{fig.dmft}. At half-filling, it is
expected that the three-site unit cell (see Fig.~\ref{fig.schematic})
is sufficient to investigate the interaction-induced order parameters,
while larger clusters should be considered to capture FFLO order
appearing in the spin-imbalanced case. Further we define the s-wave
order parameter from the anomalous Green's function $F$ as
 \begin{equation}
 \Delta(r_x,r_y)=UF(r_x,r_y)(\tau \rightarrow 0^{-})\label{eq7}
 \end{equation}
 where $(r_x,r_y)$ are the positions of the sites in the unit cell and
 $\tau$ is the imaginary time. Similarly, we denote
 $n_{\sigma}(r_x,r_y)=G(r_x,r_y,\sigma)(\tau\rightarrow 0^-)$, where
 $G$ is the normal Green's function and $\sigma$ is the spin degree of
 freedom, and define the spin-polarization as
\begin{equation}
 n_s(r_x,r_y)=  n_\uparrow(r_x,r_y)-n_\downarrow(r_x,r_y).\label{eq8}
\end{equation}
The two dimensional profile distribution of the
s-wave order parameter, $\Delta(r_x,r_y)$, and spin-polarization,
$n_s(r_x,r_y)=n_{\uparrow}(r_x,r_y)-n_{\downarrow}(r_x,r_y)$, in the
Lieb lattice are shown in Fig.~\ref{fig.dmft}. In the figure, the
18-site cluster is stacked in the
$y$-direction. The modulations of the order parameter and
spin-polarization are a clear indication of an FFLO state.

In the kagome lattice, calculations were performed on the three-site
unit cell shown in Fig.~1 in the main text. The Fulde-Ferrell (FF) ansatz
$\Delta_{{\bf j} \alpha} = \Delta_{\alpha}e^{i{\bf q}\cdot{\bf j}}$ is
included by performing the transformation $\psi_{{\bf
    j}\alpha\uparrow}\rightarrow\psi_{{\bf j}\alpha\uparrow}e^{-i{\bf
    q}\cdot {\bf r_j}}$, where ${\bf r_j}$ is the position of the
$j$th lattice site. The dependence on the momentum ${\bf q}$ of the
Cooper pairs is then included in the hopping matrices, and the
non-interacting Green's function becomes ${\bf G}^0({\bf
  k},i\omega_n)^{-1}_{ij} = (i\omega_n+h)\delta_{ij}\sigma_0 +
\mu\delta_{i,j}\sigma_0-{\rm diag}({\bf T}({\bf k - q})_{ij},-{\bf
  T}({\bf k})_{ij})$, where ${\bf T}({\bf k})$ is again
the Fourier transform of the hopping matrix. Like in the Lieb lattice,
the self-energy is assumed local, but can be different for the three
sites in the unit cell. 

The computation for the kagome lattice is performed at different
amplitudes of ${\bf q}$, with the direction fixed perpendicular to one
of the lattice vectors, corresponding to the most favorable direction
found in mean-field calculations. The chemical potentials are tuned to
achieve the same filling fractions for all different ${\bf q}$, and
the most favorable amplitude is determined by comparing the total
energies. The results for lattice filling fractions
$n_{\uparrow}\approx 2.06$ and $n_{\downarrow}\approx 1.62$ with
interaction $U=-4J$, are shown in Fig.~\ref{fig.dmft}. The computation
converged to a state with finite order parameters around
$q\approx 1.0$, indicating an FF state. The FF state at
$q\approx 0.73$ gave the lowest total energy. At these filling
fractions, the majority component has reached the flat band, so these
results confirm the mean-field observation that the FF state can exist
near the flat-band singularity.

\begin{figure}
\includegraphics[width=.99\columnwidth]{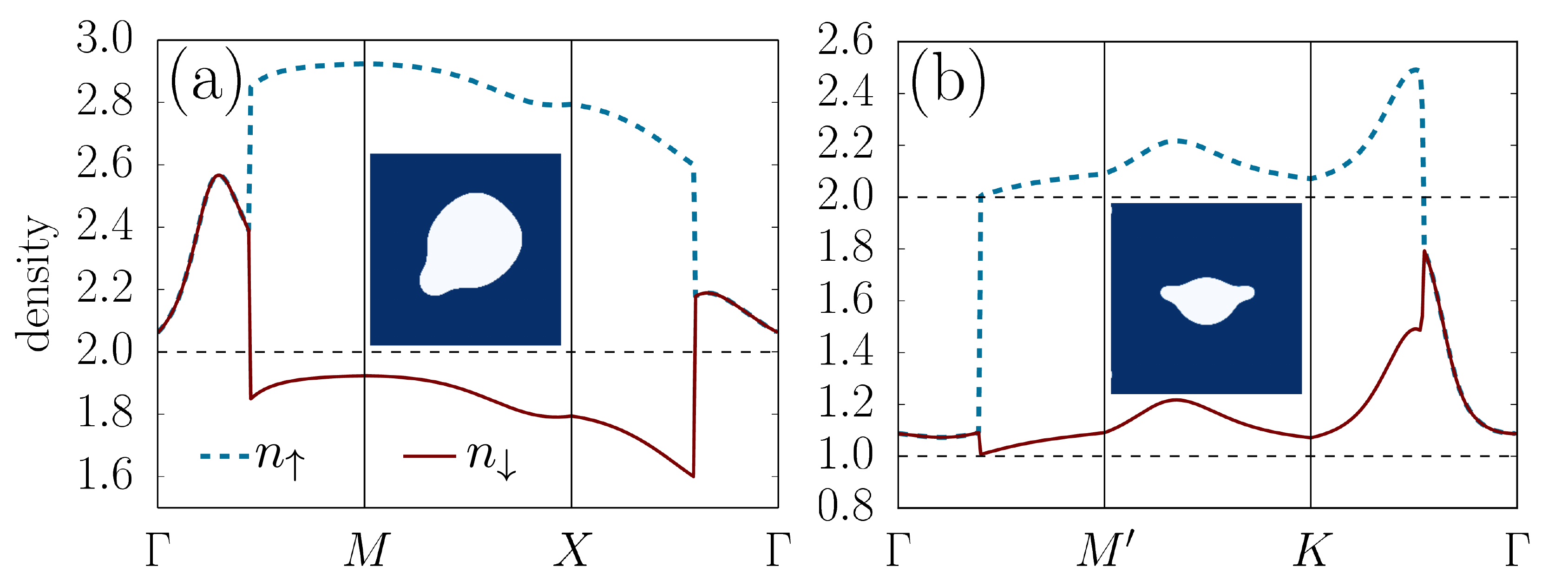}
\caption{Densities along high symmetry symmetry lines in (a) the Lieb
  and (b) the kagome lattice. Coinciding densities indicate complete
  pairing, whereas a jump in density
  $n_{{\bf k}\up}-n_{{\bf k}\down} = 1$ arises due to the presence of
  a normal gas characteristic for spin-imbalanced superfluids. The
  region where this unpaired component resides is shown as
  insets. Dark blue (white) corresponds to a value 
  of one (zero) of $n_{{\bf k}\up}-n_{{\bf k}\down}$. }
\label{fig.cross_section}
\end{figure}

\begin{figure}[t]
\includegraphics[width = .98\columnwidth]{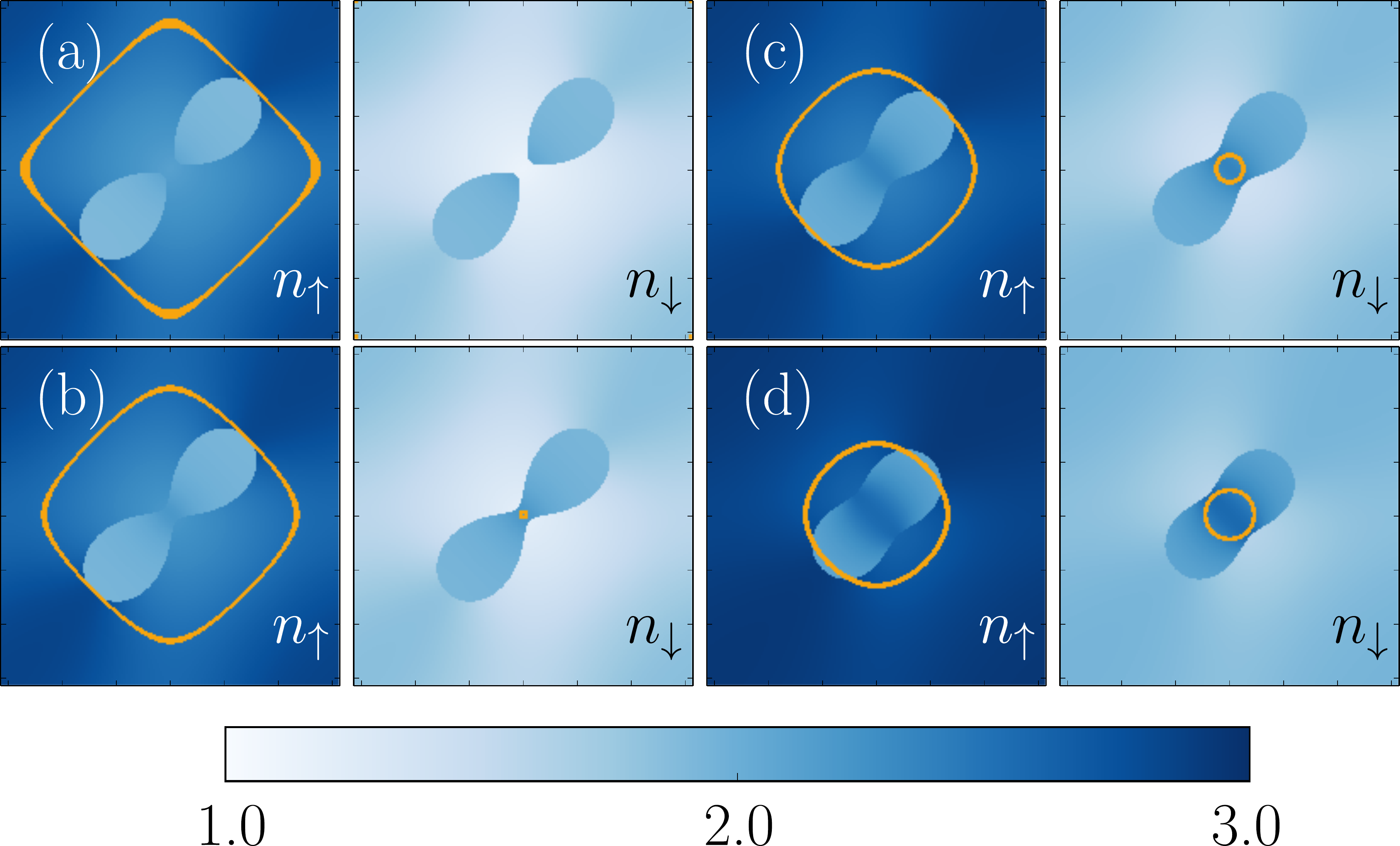}
\caption{(a) Total densities of each spin component for four different
  parameters. The shape of the deformed Fermi surface is revealed by
  the distribution of the normal part, as it changes within the
  $\eta$ phase as we move away from the crossing of the flat band and the Van Hove
  singularity, i.e. along the line of $h = 1.05\, J$; Subsequent
  panels are for (a) $\mu = J$, (b) $\mu = 1.1\, J$, (c)
  $\mu = 1.3\, J$, (d) $\mu = 1.5\, J$. Since the I-DB is completely
  filled, the color scale was truncated to the range from 1 to 3. 
}
\label{fig.Slieb}
\end{figure}

\begin{figure}
\centering
\includegraphics[width=.99\columnwidth]{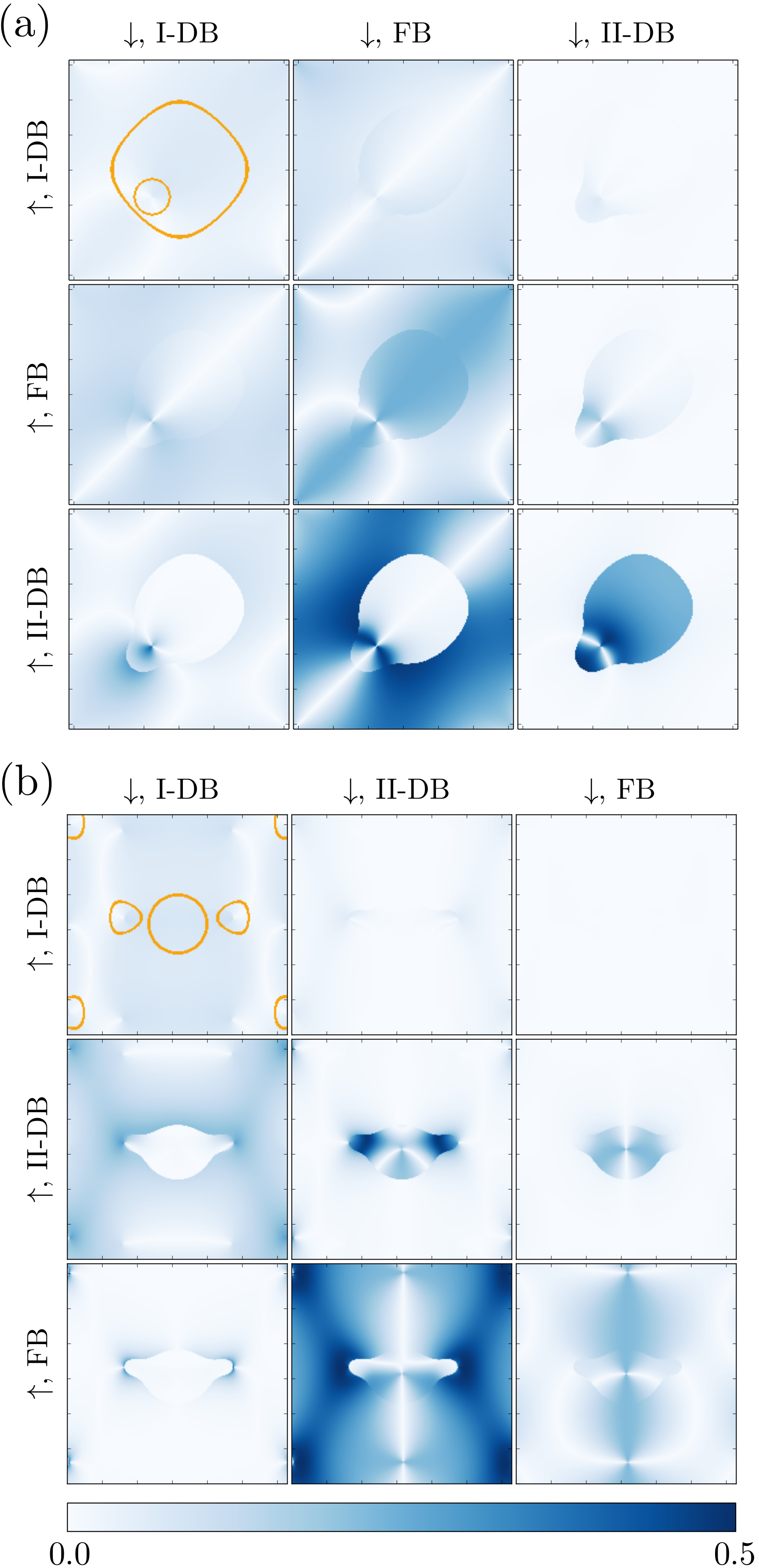}
\caption{Pairing $\langle d_{{\bf k} n \up} d_{({\bf q} - {\bf k})   m \down} \rangle$ between different bands in (a) the Lieb lattice 
  and (b) the kagome lattice at the parameters used in Fig.~\ref{fig.band_densities}. Most pairing takes place as intraband pairing in the II-DB and as interband pairing between the FB and the II-DB.}
\label{fig.pairing}
\end{figure}
\begin{figure}
\centering
\includegraphics[width=.99\columnwidth]{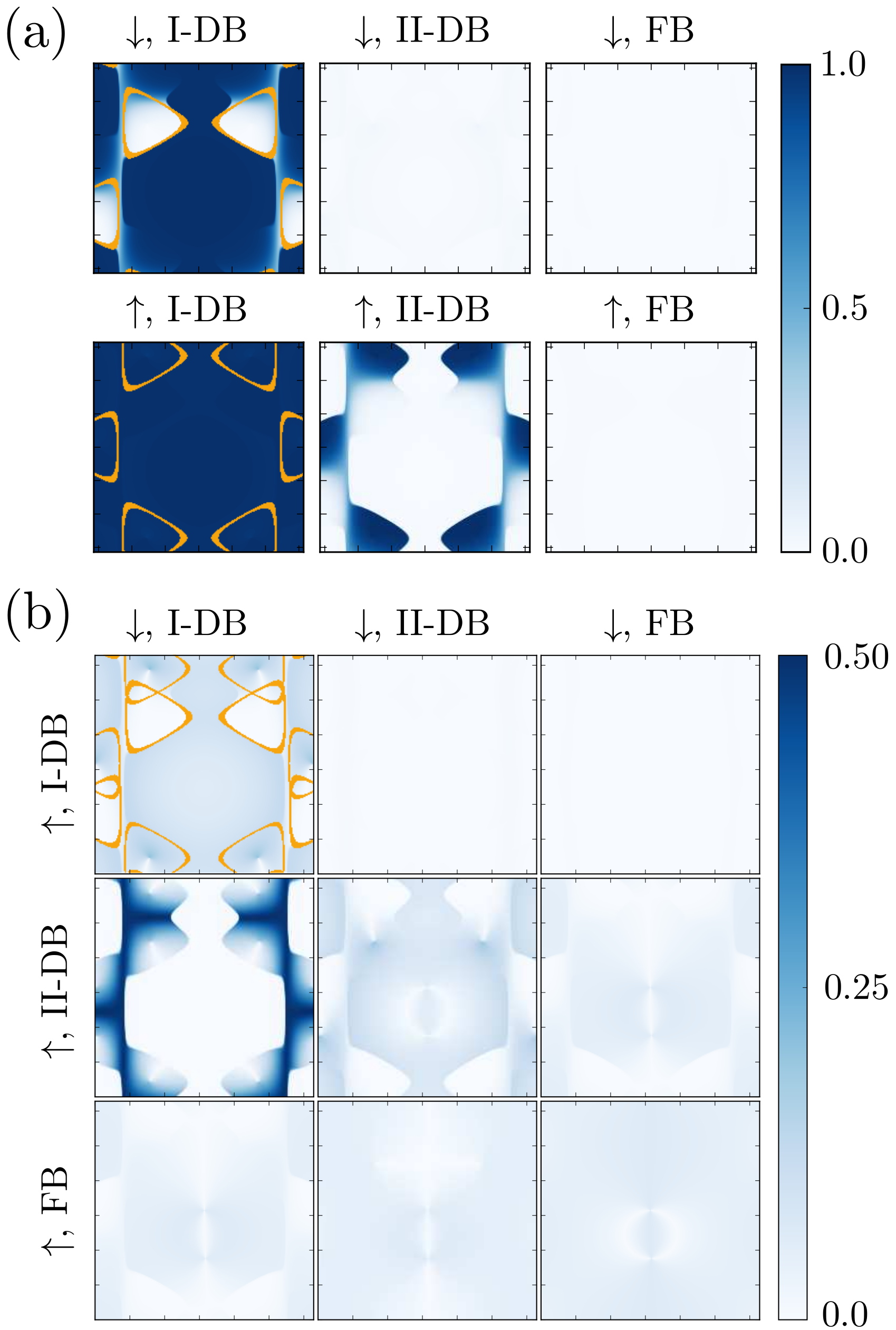}
\caption{(a) Band-resolved density profiles and (b) pairing
$\langle d_{{\bf k} n \up} d_{({\bf q} - {\bf k})m \down} \rangle$ in
the kagome lattice at $\mu=-1.0$, $h=0.9$, $U=-4J$
and $T = 0$. The pairing is mainly interband pairing between the two
dispersive bands, and correlations are most pronounced where the Fermi
surfaces of the two components are matched.}
\label{fig.other_FF}
\end{figure}

\section{Pairing mechanism}
In order to get an insight into the mechanism of pairing in these
multiband systems, we look at the band resolved densities,
$n_{{\bf k}n \sigma} = \langle d_{{\bf k} n \sigma}^\dag d_{{\bf k} n
  \sigma}^{\vphantom{\dag}} \rangle$,
where $n$ is the band index, that is, densities of each spin component
decomposed in the band basis of the single-particle Hamiltonian. As
presented in the schematic in Fig.~\ref{fig.band_densities}(a), we
find that the FS of the minority component gets shifted by a vector
$\q$ towards the Fermi surface of the majority component where the
pairing takes place --- this is the conventional mechanism behind the
FFLO state~\cite{Fulde1964,Larkin1965,Kinnunen2017}. In a square
lattice, this leads to nesting which stabilizes the FFLO
state~\cite{Koponen2007,Kinnunen2017}. In our case this is {\it
  intra}-band pairing, i.e. pairing between atoms from the same band,
as will be explained later. The calculated band-resolved densities are
shown in Fig.~\ref{fig.band_densities}(b) for the Lieb lattice and in
Fig.~\ref{fig.band_densities}(c) for the kagome lattice. The lower
dispersive band (I-DB) remains almost completely filled (and therefore
we do not plot it), while deformation of the density distributions
takes place in the upper dispersive band (II-DB) in the region where
the FSs match.

An interesting effect can be observed for atoms residing in the
FB. For one component they remain completely unaffected, 
while for the other the distribution of atoms, which was
initially flat, gets deformed in such a way as to mimic the density of
the first component in the II-DB. In the case of a
Lieb lattice (kagome lattice) the FB remains completely filled
(completely empty) for the majority (minority) component, while for
the minority (majority) component the distribution of atoms gets
deformed. This suggests an {\it inter}-band pairing between the atoms
in the FB and atoms in the II-DB. This is an
energetically favorable process, as the atoms in a FB can
rearrange at vanishing energy cost due to flat dispersion
relation. Such density rearrangement without energy cost is the key
physical role of the FB in enhancing exotic pairing.

The excess atoms of the majority component, that do not take part in
the pairing, form a normal gas. Its presence can be seen in the total
density traced along the high-symmetry lines, as well as in the
differences $n_{{\bf k}\up}-n_{{\bf k}\down}$, as shown in
Fig.~\ref{fig.cross_section} for both lattices. The density profiles
of the paired components are matched up to a shift by a constant; for
some momenta $\bf k$ there is a jump in the densities of the two
components. This is due to the presence of a normal gas. Since, as is
stated by Luttinger's theorem, the number of available states inside
the Fermi sphere does not change upon interactions, this constant
shift is $n_{\bf k \up} - n_{\bf k \down} = 1$, see
Ref.~\cite{Kinnunen2017}. This mechanism can be seen also in the
band-resolved densities in Figs.~\ref{fig.band_densities}(a), (b). The
presence of the normal gas in the upper dispersive band gives rise to
an observable FS seen as sharp density jumps. Even though the normal
component does not participate in the pairing, its Fermi surface is
deformed by the pairing mechanism of the other
atoms.

As we approach the flat band singularity within the $\eta$ phase in
the phase diagram of the Lieb lattice, the deformation of the Fermi
surface becomes more and more pronounced. This deformation is such
that there be as large a matching as possible between the two
FSs. That is where most of the intraband pairing takes place. In
Fig.~\ref{fig.Slieb} we show four examples of cumulative density for
each spin component along the line of $h = 1.05\, J$. When one of the
non-interacting FSs vanishes at the Dirac point, the deformation is the
most dramatic, and the continuity of the FS is broken.

To gain further understanding of the
nature of pairing, we study pairing
correlations between different bands, $\langle d_{\vphantom{()}{\bf k} n
  \up} d_{({\bf q} - {\bf k})   m \down} \rangle$, where
$n$ and $m$ are band indices. As can be seen in
Figs.~\ref{fig.pairing}(a),(b), the lattices feature both intra- and
interband pairing. Intraband pairing occurs mostly between particles
on II-DB, and is most pronounced in the region where the Fermi
surfaces match. This is similar to what is 
found in the square lattice, where particles on the same energy band
can pair due to the shift of one FS by $\q$. The Fermi
surface of the normal component is reflected also in the pairing
correlations, and intraband pairing within II-DB is completely absent
in the region where the unpaired particles reside.  The other
prominent pairing is between particles on the FB and those on
II-DB. Again, the FS of the normal component is visible as
sharp jumps between low and high correlations. Contrary to intraband
pairing, this interband pairing occurs mostly where the unpaired gas
lies, and the paired components occupy different energy
bands. Pairing is made possible in this situation by the possibility
of atoms on the FB to readjust their density profile to mimic that of
the other component on II-DB at low energy cost. 

Correlations between other bands, albeit smaller, are also present. In
particular, also particles of the majority (minority) component on the
FB contribute to pairing in the Lieb (kagome) lattice. Moreover, the
various pairings give further indication that the unpaired particles
are distributed among different bands. 

To better understand the effect of the flat band, it is instructive to
compare the pairing mechanisms in the FF phase near the FB singularity
to those in the other FF region found for the kagome
lattice. As can be seen from the band-resolved densities and
correlations shown in Fig.~\ref{fig.other_FF}, the FB is almost empty
for both components, and contributes little to the
pairing. The dominant pairing is interband
between atoms on the first and second dispersive bands.
Interestingly, even though the Fermi surfaces are perfectly
matched at zero $\q$, the FF phase is favorable. This is due to the
different distributions of the components: the minority component
occupies the center of the BZ, whereas the majority
component occupies the corners. The momentum $\q$ allows for the Fermi
seas of the two components to overlap slightly, increasing the number
of states near the Fermi surface that can pair.

The comparison with pairing correlations near the FB highlights the
effect of a FB on the pairing mechanism. Intraband pairing is almost
absent in the FF region away from the flat band, whereas both intra-
and interband pairings are found near the FB singularity. Moreover,
the possibility for atoms on the FB to rearrange allows for pairing to
occur in a large region of the BZ, instead of being
limited to the comparatively small region where Fermi surfaces are matched.

The pairing correlations in the band basis for the $\eta$ phase at the flat band (near the point where the singularities cross) show the same mechanism as described for the generic FF phase: in the Lieb lattice intraband pairing is mostly concentrated within the II-DB and within the flat band, and the interband pairing between the flat band and the II-DB. The difference is that the deformed FS in the $\eta$ phase is symmetric with respect to the $\Gamma$ point; this is due to the four-fold symmetry of the original, non-interacting FSs. 

\section{Experimental prospects}

While other possibilities also
exist~\cite{Slot2017,Drost2017,PhysRevLett.114.245504,PhysRevLett.114.245503,PhysRevB.93.054116},
ultracold quantum gases may offer the most immediate realization of
our predictions. Lieb and kagome geometries have already been realized
by optical
lattices~\cite{PhysRevLett.108.045305,Taiee1500854,PhysRevLett.118.175301}
and novel techniques such as digital mirror devices and
holograms~\cite{Hueck2017,Gauthier2016,Zupancic2016} allow further
flexibility. Our mean-field calculations give critical temperatures
$k_B T_c$ from around $0.2\, J$ to $0.5\, J$~\cite{TylutkiNew}. In 2D,
the Berezinskii-Kosterlitz-Thouless (BKT) temperature for
superfluidity is typically smaller than the BCS one but can be of the
same order of
magnitude~\cite{Peotta2015,Julku2016,Liang2017}. Deformations and
nontrivial pairing correlations may appear in these flat band systems
already well above the critical temperature, which is an interesting
topic of future study.

\section{Conclusions}

In summary, we studied the attractive Hubbard model on the Lieb and
kagome lattices, both featuring a FB. We found a stable FFLO phase,
present due to inter- and intraband pairings that involve the FB. This
mechanism of spin-imbalanced pairing relies on complete deformation of
the density of one pairing component, enabled by the FB, and is
therefore strikingly different from the conventional minority FS shift
(and nesting in lattices). Flat band singularities are known to
enhance magnetism \cite{Mielke1992,Tasaki1992,Mielke1993} and
superfluidity~\cite{Khodel1990,Khodel1994,Heikkila2011,Kopnin2011,Liang2017,Peotta2015,Julku2016,Yamamoto2013};
here we have shown that, in the case of spin-imbalanced pairing, not
only does it enhance interactions, but also it makes the pairing
mechanism qualitatively different. Since experimental preparation of
artificial lattice quantum systems is advancing
rapidly~\cite{Torma2015,Kuhr2016,Drost2017}, our predictions may show
the route to direct observation of spin-imbalanced pairing and
superfluidity.

\section{Acknowledgments}
This work was supported by the Academy of Finland through its Centres
of Excellence Programme (2012–-2017) and under project NOs. 284621,
303351 and 307419, and by the European Research Council
(ERC-2013-AdG-340748-CODE). This project has received funding from the
European Union’s Horizon 2020 research and innovation programme under
the Marie Sk{\l}odowska-Curie Grant Agreement No. 702281
(FLATOPS). T.I.V. acknowledges support from the V\"{a}is\"{a}l\"{a}
foundation. Computing resources were provided by CSC -- the Finnish IT
Centre for Science and the Triton cluster at Aalto University.

\end{document}